\documentclass[
twocolumn,
]{ceurart}

\usepackage{listings}
\usepackage[ruled, vlined]{algorithm2e}
\usepackage{algpseudocode}

\begin{document}

\copyrightyear{2022}
\copyrightclause{Copyright 2022 for this paper by its authors. Use permitted under Creative Commons License Attribution 4.0 International (CC BY 4.0)}

\conference{The IJCAI-ECAI-22 Workshop on Artificial Intelligence Safety (AISafety 2022),
  July 24--25, 2022, Vienna, Austria}

\title{Let it RAIN for Social Good}


\author[1]{Mattias Brännström}[%
orcid=0000-0003-3113-2631,
email=mattias.brannstrom@umu.se,
]
\cormark[1]
\address[1]{Umeå University,
  Universitetstorget 4, 901 87 Umeå, Sweden}

\author[1]{Andreas Theodorou}[%
orcid=0000-0001-9499-1535,
email=andreas.theodorou@umu.se,
url=http://www.recklesscoding.com]

\author[1]{Virginia Dignum}[%
orcid=0000-0001-7409-5813,
email=virginia.dignum@umu.se
]

\cortext[1]{Corresponding author.}

\begin{abstract}
Artificial Intelligence (AI) as a highly transformative technology take on a special role as both an enabler and a threat to UN Sustainable Development Goals (SDGs). AI Ethics and emerging high-level policy efforts stand at the pivot point between these outcomes but is barred from effect due the abstraction gap between high-level values and responsible action. In this paper the Responsible  Norms (RAIN) framework is presented, bridging this gap thereby enabling effective high-level control of AI impact. With effective and operationalized AI Ethics, AI technologies can be directed towards global sustainable development.
\end{abstract}


\maketitle

\section{Introduction} 

Several recent and comprehensive reviews make clear that there is a strong connection between large-scale change and developments in Artificial Intelligence (AI) \cite{vinuesa,castro}. All of the 17 Sustainable Development Goals (SDGs) for Sustainable Development are believed to be moderately or strongly affected by AI technology. Studies show that 59 of the sustainable development targets might actually be inhibited by AI and there is reason to believe this is a low estimate \cite{vinuesa}. The large scale predicted effects of AI take up a complicated role as some progress towards sustainability might be \emph{dependent} on AI for the required changes. Some studies even go as far as to term this technological progression a ``vector of hope'' \cite{castro}. 
Research gaps exist regarding the large scale effects in the interplay between AI technologies and society where AI related change could instead exacerbate negative narratives and global inequalities \cite{sartori2022sociotechnical,castro,Sampath}. 

A central role in determining the outcome, positive or negative, of AI on large-scale sustainability and the SDGs is taken by AI Ethics. It is widely recognized that effective soft and hard policies on AI technologies are needed to ensure positive outcomes. Many attempts at high-level soft policy already exists by intergovernmental organisations, e.g. the European Commission's ``Guidelines for Trustworthy AI'' (GTAI), but also by professional bodies, e.g. IEEE. Such policy documents focus on advocating high-level ethical principles such as \emph{fairness}, \emph{transparency}, \emph{accountability}, and \emph{respect for human values} \cite{theodorou2020towards}. 

An often mentioned problem of the high-level guidelines is that they are at times too abstract to be applied to any particular case and and at other times too specific by mentioning problems which might not exist in a particular application. There is no particular level of abstraction that solves this problem for high-level policy as guidelines either become too abstract or too extensive. A gap thus appears between high-level policy and any practical application 
\cite{theodorou2020towards,mittelstadt2019principles}.

Further exacerbating the problem is that the socio-technical domain typically consist of not a single actor, the AI developer, but an interplay between developers, procurers, customers and users \cite{rubenstein2021acquiring,falco2021governing}. It is within this socio-technical multi-actor sphere where the effects of AI on society develop \cite{rubenstein2021acquiring,falco2021governing,mittelstadt2019principles}. Understanding this interplay and successful bridging this \emph{abstraction gap} between high-level policy and particular application is of central importance in establishing socially-beneficial AI.


The abstraction gap is not only a problem from a regulatory perspective.
For the individual developer, procurer, or any  other actor dealing with emerging AI applications where the gap severs the link between design and organisational choices on one hand and outcomes, ethical or otherwise, on the other. As there is no clear link between the particularities of an AI application and high-level ethical goals, there is no clear path forward even for actors on all levels who desire to act responsibly. 

Currently, bridging this gap require expert involvement and analysis. This contribute to increase the divides and inequalities already present in society, decrease the transparency of AI Ethics itself and undermine trust in AI technologies. In other words, negatively contributing towards the SDGs. Effects like these are even more prominent in areas where both expertise and effective governance structures with a strong ethical focus is lacking; leading to an AI ethical void in the most sensitive areas for increasing global inequality \cite{Sampath}.

The solution to bridging the gap is context awareness and structure, in policies, tools, assessment procedures, and in communicating that context across actors. AI Ethics without the specific context lacks solutions and low-level technical approaches loses sight of the goals and the larger effects \cite{mittelstadt2019principles,theodorou2020towards,bird2020fairlearn}. A continual chain encapsulating all levels of the socio-technical landscape is needed to ensure relevance to both actual applications and the larger society \cite{falco2021governing,rubenstein2021acquiring}. When such a chain is explicit, it can drive the transformative effects of these emerging technologies towards sustainable change in line with the SDGs.

In this paper, a solution for bridging the abstraction gap is presented: the \textit{Responsible AI Norms} (RAIN) framework. RAIN breaks down abstract high-level policies into actionable norms by connecting them with socio-technical contexts in a structured way. 
The formal structure of RAIN provides clarity in the connections between policies and actual AI applications on all levels and thereby enables effective policy-making and policy compliance with low overhead. The formal specifications also enable reproducibility and auditability of the RAIN-produced requirements.

The paper is structured as follows: First, a brief theoretical background is provided. Then, RAIN is described in detail using examples (given in italics). Finally, the paper concludes with a discussion how RAIN aids the transition towards the SDGs on all levels as well as directions for future work.

\section{Background} 

    Value Sensitive Design (VSD) is a methodology for centering design around abstract high-level values, by embedding values into the socio-technical context where they are being used, value-conflicts and key concrete design requirements can be identified \cite{vsd,van2013translating}. VSD starts by placing the focus on a socio-ethical value relevant for the use case at hand. From this perspective, any associated values, stakeholders, and technologies are found by iterative exploration. Harms and benefits of each identified group of stakeholders are determined and connected to relevant values and values are prioritized. After this mapping has taken place, conflicts between values, technological solutions and project goals can be brought to forefront in the design process. Key here is the exploration of how a value impacts design by exploring the intersections between value, stakeholders and technological use case. The goal of the process is to facilitate discussion and understanding.

    The VSD process can be made more structured using the \emph{count-as} operator to break down high-level norms into contextualised lower-level norms\cite{van2013translating}. Linguistically counts-as represents the construct `X counts as Y in context Z', and has been well described formally in depth in \cite{grossi2007designing}. Counts-as enables the expression of values in specific contexts as sub-norms finally connecting to concrete design choices.
    
    Building upon VSD is the Glass-Box framework \cite{aler2019glass}. The Glass-Box approach demonstrate how the the same procedure can be used to to retrieve testable requirements. The Glass Box consists of two phases which inform each other: interpretation and observation. The interpretation stage translate values into specific design requirements by using the VSD approach where the relationship between values, norms, and requirements can be formally represented using \emph{counts-as} and modal logic \cite{aler2019glass,grossi2007designing}. 
    
    Once found, the low-level requirements inform the observation stage of the approach. The requirements, now linked to high-level values, can be automatically or even continually assessed to determine to which degree a solution fulfills its stated values.
    However, two key limitations remains: The interpretation step must, as VSD, be done separately for each AI application, something that requires significant expertise and may produce diverse results. It also focus on measurable requirements, which limits the approach to the technical compliance while AI is a socio-technical system.

\section{RAIN} 

    The RAIN framework provides further structure compared to VSD in the hierarchical breakdown process of values in a way that makes the resulting norms hierarchy with its context-sensitive requirements reusable. The RAIN norms hierarchy is also just as apt for questionnaire-type assessment questions as automatic tests or continual monitoring, and, thus, extending the use case of the Glass-Box to the socio-technical sphere.
    
    A reusable norms hierarchy go a long way to reduce the overhead of Ethical AI. It shifts the focus towards application features which is more readily dealt with by current technical B2B-landscape. It also enables a structured way to work with and communicate around policy and AI Ethics between societal actors. In addition the RAIN framework also serve as a kind of knowledge elicitation from experts. This embedded knowledge can be of aid in settings where such expertise might not be available. Such clarity around tangible impacts, responsibility and concrete ethical choices is key for transparent and accountable use of AI technology which drives towards the SDGs rather than inequality and exploitation.

        
    In this section the RAIN framework will be described in detail, starting with the creation of the norms hierarchy, then exploring the connection to socio-technical context, scoring mechanisms and, finally how to derive assessment results and projections of assessments on particular policies.
    
    
    \subsection{Overview of the RAIN Pipeline}
    
        The RAIN framework can be seen as consisting of four fundamental components. The heart of the framework is the \textit{RAIN Graph}. The RAIN Graph contains a structured and contextualized norms hierarchy. The section below will detail it’s methods of construction from an existing policy. 
        Building around the Graph is a three part pipeline starting with the \textit{context layer}, which captures the context features of a particular AI application in order to determine which contexts of the RAIN Graph which are active. Having established this, the \textit{assessment layer} can be used to determine compliance to identified norms. Finally, the \textit{results layer} concerns aggregation and extraction of results from the Graph and Assessment.
        
    \subsection{Building the RAIN Graph} 
    
        The RAIN Graph captures AI policy in a structured manner.
        In this section it will be described how such a Graph can be derived from a high-level policy but also expanded to particular contexts not explicitly mentioned in such a policy. High-Level Policy (HLP) will in this section be defined as any policy, guideline, standard or the like which primarily bases itself upon High-Level ethical Values (HLV) and presents challenges to these from the use of AI technology, solutions to such challenges, or requirements on action to alleviate such challenges. These challenges, regardless of which form they appear will be termed AI Issues.
        
        The framework description will be aided by standard \textit{Description Logics} extended with the context scope ($x:y$, where $y$ applies in context $x$), counts-as ($\Rightarrow_c$) operators and context relation $\preceq$ as described by \cite{grossi2007designing}. The formalism make relationships exact and explicit, something which is required for the framework to work in reproducible inter-operability and communication of  concerns between actors. The formalisation also lends itself readily to implementation.  
        
        \subsubsection{A scaffold for High-level AI Policy}
        
        Before breaking down and structuring any AI policy, we start by defining a simple scaffold in which to understand them. We specify that:
        \begin{align}
            T_c : & HLV \sqsubseteq AI Ethics \\
            T_c : & ethical AI \equiv AI \sqcap \neg \exists violate.AI Ethics \\
            T_c : & issue \equiv \exists violate.AI Ethics
        \end{align}
        That is, in the top context HLV are sub-concepts of \textit{AI Ethics} (1). \textit{Ethical AI} is AI which does not violate \textit{AI} Ethics (2) and an \textit{issue} is something that do (3).
        
        Given this scaffold, particular a policy can be seen as sub-context providing additional detail primarily to the abstract concepts \textit{HLV} and \textit{issue} by specifying sub-classes. While policies are not typically written in such a structured manner we can use this scaffold to frame the content and see them as sets of statements about \textit{HLV} and sets of statements about \textit{issue}s. The following parts of this framework will help to extract detail so framed. Policies also frequently mention particular technical features or stakeholders, if so these too are seen as content of the policy.
        
        \textit{Ex. GTAI presents several issues around the HLV Privacy which is expressed as consisting of Right to Privacy, Right to Data Protection, and Data Governance. Among issues are use of personal data in training and use of transmission and storage of personal data, all of which \emph{violate} subsets of \emph{Privacy} and thus \emph{AI Ethics}. Some of these concerns are not in the provided assessment-questions but in the descriptive text.}
        
        \subsubsection{The RAIN scaffold and scoring model}
        
        With the scaffold for the policies in place, we can follow this with a new context $R_0 \preceq T_c$ forming the basis of the rain framework. We can describe $R_0$ as:
        \begin{align}
            R_0 : & value \sqsubseteq HLV \\
            R_0 : & violate{\text -}1 \sqsubseteq violate \\
            R_0 : & violate{\text -}2 \sqsubseteq violate{\text -}1 \nonumber \\ 
            R_0 : & violate{\text -}3 \sqsubseteq violate{\text -}2 \nonumber
        \end{align} 
        In the practical applications of the RAIN framework, a graded scoring model of maturity levels is used: 1 implies that the system violate the high-level requirements to a minor degree and each other level indicate lesser degrees of compliance with more serious violations. Each level is given a concrete definition as to what type of requirements it contains. Possible attached meaning to the scoring model is not the focus of this paper. Hence, we use a 3-tiered score model that will be used as an example how scoring mechanisms are tied to the framework (4,5). Different numbers of levels and different definitions of each level work in the same way. 
        
        The scoring model here described results in a threshold model of aggregation. Within each category the aggregated score will be the worst score within that category. This approach counteracts `ethics washing' the approach of doing something less-relevant well to make up for major failures in more relevant areas. It also helps to highlight the ethical issues with an application where the most difference can be made.
        
        \subsubsection{The RAIN Graph}
        
        A RAIN Graph, $G$ can be described to contain the following concepts
        \begin{itemize}
            \item \textbf{value} A parsimonious sub concept of HLV. Often values in policy are expressed using several sub values or norms, these are here separated. We will term the set of all \textit{value}s $v \in G$ as $V$
            \item \textbf{stakeholder} Reflecting a perspective of concern for a stakeholder group. We will term the set of all \textit{stakeholder}s $s \in G$ as $S$.
            \item \textbf{socio-technical feature} A socio-technical use of technology. We will term the set of all socio-technical features $f \in G$ as $F$.
            
            \item \textbf{RAI norms and contexts} A norm $n(s,f) \in G$ represent a particular challenge caused by some socio-technical feature $f \in F$ to a value $v \in V$ with respect to a stakeholder concern $s \in S$. We will term the set of all such norms $n$ as $N$.
            Every such \textit{norm} $n \in G$ will be embedded in a sub-context $n_c \preceq R_0$ such that $n_c : n \sqsubseteq (f \sqcap v \sqcap s$).   
               
        \end{itemize}

        
        The sets  $V, S, F, N, N_c$ are considered to be holding semantically distinct items. For the algorithms~1~and~2, we define the operation \textbf{merge} to mean an addition that preserves semantic distinctness. In the case of RAI norms, $N$ multiple distinct issues with corresponding assessment lists can have the same semantics but will be distinct if assessment criteria are taken into account. If so they occupy the same norms context as they are activated by the same features.

        \subsubsection{RAI Norms and contexts} 
        
        The RAI \emph{norms} are the central content of the RAIN Graph. These norms can be seen as representing the junction between a value, a subject, and a circumstance. Or value, subject, and action. Through these norms, it is possible to determine what features of a given context which are related to which values and for whom. These relationships are the main purpose of the RAIN Graph and how it helps to bridge the abstraction gap.
        
        Since each of the RAIN nodes identifies a particular threat, it can be accompanied with a corresponding set of requirements alleviating that threat. In this manner, a context-sensitive assessment of how a given AI application complies with one or several policies can be expressed as the degree of which it fulfills the requirements selected by its features. Some types of the technical requirements can be verified in an automated manner; in other words, the RAIN Graph fulfills the interpretation stage of the Glass Box by identifying in which ways it is relevant to monitor an application with regards to ethical concerns. Other requirements, concerning organisational features, design choices, or documentation, require a wider socio- intervention by stakeholders. Such requirements instead lend themselves to manual assessment procedures.
        Formally we can represent these RAI norms and their accompanying assessment rules as their own contexts $n_c \preceq R_0$ where $n_c$ represents the active presence of a stakeholder and feature instance in the context of the application. The general structure of this context also including the foundation of the assessment layer can be expressed as follows:
        \begin{align}
            n_c : & N_a \Rightarrow_c (v \wedge s \wedge f) \\
            n_c : & Assessment{\text -}1 \equiv \exists violate{\text -}1.N_a \\
            n_c : & Assessment{\text -}2 \equiv \exists violate{\text -}2.N_a \nonumber \\
            n_c : & Assessment{\text -}3 \equiv \exists violate{\text -}3.N_a \nonumber
        \end{align}
        
        \subsubsection{Operational semantics algorithms}
            
            The RAIN Decomposition Algorithm encodes a policy into the graph. A second algorithm described here, the RAIN Expansion Algorithm, fills out the missing areas of concern and expands the policy with consideration of a potentially new area of socio-technical context.
            
            Algorithm \ref{DA}, the decomposition algorithm or backwards algorithm goes from policy and provides a RAIN graph encoding of its content.
            \begin{algorithm}[t]
            \caption{RAIN Decomposition Algorithm\label{DA}}
            \DontPrintSemicolon
            \KwData{$\textbf{P}$, a policy\;}
            \KwData{$G(V,S,F, N, N_c)$, a RAIN Graph\;}
            \Begin{
                \For{\textit{hlv} $\sqsubseteq$ HLV $\in$ $\textbf{P}$}{
                    \textbf{merge} component \textit{values} $v$ of \textit{hlv} to $V$\;
                }
                \If{Explicit stakeholders $\in \textbf{P}$}{
                    \textbf{merge} component stakeholder concerns $s$ of policy into $S$\;
                }
                \If{Explicit socio-technical features $\in \textbf{P}$}{
                    \textbf{merge} component stakeholder concerns $s$ of policy into $S$\;
                }
                \For{$i \sqsubseteq \textit{issue} \in \textbf{P}$}{
                    $V_i \gets$ Values $v \subset V$ impacted by \textit{issue} $i$\;
                    $S_i \gets$ Stakeholder concerns impacted by \textit{issue} $i$\;
                    $F_i \gets$ Socio-technical features which \emph{must} be present for \textit{issue} $i$ to threaten $V_i$ with regards to $S_i$\;
                    \textbf{merge} concerns $S_i$ into $S$\;
                    \textbf{merge} features $F_i$ into $F$\;
                    
                    \textbf{merge} norm $n(V_i, S_i, F_i)$ into $N$ and $N_c$\;
                }
            }
            \end{algorithm}
            
            \begin{enumerate}
                \item Start with a policy document. \textit{Ex. GTAI}.
                \item Identify the top values which are directly impacted or taken into consideration by the policy. \textit{Ex. Privacy}.   
                \item Consider the characterization of each of these high level values in order to break down each of these high-level concepts into singular areas of concern. The key here is to be parsimonious. One concern or concept per item.
                
                \textit{Ex: {\text{Right to Privacy, Data protection, Data Governance}} $\sqsubset$ Privacy $\in$ GTAI}
                \item \textbf{merge} the resulting derived top values or top norms form the \textit{value}s of the Graph with respect to this policy.
                \item If Stakeholders or particular socio-technical features are explicitly mentioned in the policy, repeat the previous step for them as well in the same manner. \textit{Ex. \emph{End Users} and \emph{Developer} are mentioned stakeholders in GTAI.} 
                \item Go through each Issue raised by this policy and state it in connection to at least one Value and at least one Stakeholder. In addition determine what socio-technical feature which must be in place for the issue to exist. It might be a way of dealing with data, a particular technology, or a particular use case, for example. \textit{Ex: \emph{Handling of the personal data} of \emph{End Users} is required for issues of \emph{GDPR} in GTAI.}
                \item The identified issue or problem can now be stated as one or more RAI norms which state that this Feature threaten the identified Value with respect to the identified Stakeholder. These RAI norms are added to the Graph in their own context, as described above. 
                
                \textit{Ex. personal data$_c$  $: N_pd \Rightarrow_c ($Personal data $\wedge$ End User $\wedge$ Data Governance$)$.}
                \item When all the issues mentioned in the policy are treated in this manner one can consider the content of the Graph to contain the explicit parts of the policy. However since most policy have selected some level of abstraction and scope, it is likely that many intended issues related to AI Ethics are not yet mentioned in the Graph. These are captured using Algorithm \ref{EA}.
            \end{enumerate}
            
            Algorithm \ref{EA} consider each intersection of identified features, concerns and values in order to fill in the blanks. It can also be used for considering a particular set of socio-technical features in the light of the values and stakeholder concerns in the Graph. In this manner the Graph can be extended to cover new particular contexts.
            
            \begin{enumerate}
                \item Add any socio-technical features to be considered to the Graph. \textit{Ex. For the example in the next section, features common to home automation, voice control and human interaction such as e.g. \emph{Remote Processing} and \emph{Passive Recording}. Socio-technical features such as Vulnerable End Users are also relevant to the elderly care example below.}
                \item For each intersection between a value, a stakeholder concern and a socio-technical feature, consider the possible ways in which the value is challenged with regards to the stakeholder concern. Issues identified in this manner is treated just as in Algorithm \ref{DA} in order to add new RAI norms and RAIN-norm contexts to the Graph.
                \textit{Ex. \emph{Remote processing} interacts strongly with the GTAI values already in the graph, regarding \emph{Privacy}, \emph{Robustness} and \emph{Transparency}. Each interaction give rise to RAI Norms.}
                \item When all features are considered, features which can not be connected to both a value and a stakeholder are removed. 
                \textit{Ex. If features were added at step 1 which were of no consequence, then they are removed here.}
            \end{enumerate}
            
            \begin{algorithm}[t]
            \caption{RAIN Expansion Algorithm\label{EA}}
            \DontPrintSemicolon
            \KwData{$G(V,S,F, N)$, a RAIN Graph\;}
            \KwData{$F_{new}$ a set of new socio-technical features\;}
            \Begin{
                \textbf{merge} $F_{new}$ into $F$\;
                \For{$(f, v, s) \in F \times V \times S$}{
                    \For{Issues $i$ which $f$ threaten $v$ with respect to $s$} {
                        $V_i \gets$ Values $v \subset V$ impacted by Issue $i$\;
                        $S_i \gets$ Stakeholder concerns impacted by Issue $i$\;
                        $F_i \gets$ Socio-technical features which \emph{must} be present for Issue $i$ to threaten $V_i$ with regards to $S_i$\;
                        \textbf{merge} norm $n(V_i, S_i, F_i)$ into $N$ and $N_c$\;
                        \textbf{merge} concerns $S_i$ into $S$\;
                        \textbf{merge} features $F_i$ into $F$\;                
                    }
                }
                \If{$\emptyset = \{n | n \in N$ relates to $f\}$}{
                    \textbf{remove} $f$ from $F$ \; 
                }
            }
            
            \end{algorithm}            
        
        \subsubsection{Multiple policies and coverage}
        
        Algorithm \ref{DA} \& \ref{EA} are both idempotent and can be used repeatedly to merge several policies into a single RAIN Graph. If policies overlap in values and issues, parts of the graph might be unchanged by such additions. A special case of interest is when policies have values which are defined differently. That ethical values lack a universal definition is a common mentioned problem \cite{theodorou2020towards}. This is actually not a problem for the RAIN Graph as differing definitions mean their component \textit{values} differ. In this manner it is possible to combine even apparently conflicting policies in the same RAIN Graph. Untangling these possibly conflicting viewpoints is handled by their different semantics and the activation of different contexts, and the result projection mentioned below. 
        
        As policy are combined into the RAIN Graph in this manner it is possible to define a RAIN Graphs \emph{coverage}. Two things must pertain for the RAIN Graph to have coverage of some particular area of AI Ethics.
        \begin{itemize}
            \item A RAIN Graph have \emph{coverage} of a particular policy if merging it to the RAIN Graph using the RAIN Decomposition Algorithm (Algorithm \ref{DA}) would result in no change to the Graph.
            \item A RAIN Graph have \emph{coverage} of a particular area of socio-technical context with respect to the policy it covers if merging its Features to the RAIN Graph using the RAIN Expansion Algorithm (Algorithm \ref{EA}) would result in no change to the Graph.
        \end{itemize}
        
        \textit{Ex. The graph in the example have \emph{coverage} for GTAI, voice recognition, home automation and human interaction. The process can be repeated to add \emph{coverage} for national safety guidelines and the AI policy of local jurisdiction. Even the particular policy of a procuring organisation can be added by using Algorithm \ref{DA} and \ref{EA}}.

    \subsection{The RAIN pipeline}
        
        For the purpose of assessment, the RAIN Graph can be embedded into a pipeline with the following three steps:
        \begin{itemize}
            \item \textbf{Context layer} capturing the socio-technical context and identifying stakeholders, top level values and policies. The output of this layer is context features and activated values.
            \item \textbf{Assessment layer} providing context-specific testable requirements, satisfying the identified norms on a five-step scale of compliance.
            \item \textbf{Result layer} aggregating the result of the individual norms onto the high-level values as well as projections upon compatible policies of choice. 
        \end{itemize}
        
        The pipeline can be part of an assessment process, an iterative development process or automatic monitoring of policy compliance.
        
        For the rest of this section, a voice-controlled home-automation system will be used as a running example. A public-sector procurer is evaluating the compliance of said system against the GTAI  before its purchase and use in elderly care. This particular example is just one low-level interaction, but it such small interactions aggregate into the large scale societal effects towards or against the SDGs. The example is given in italics.

        \subsubsection{Context layer and context features} 
        
        The \textit{features} of the Graph are, as per the Algorithms \ref{DA} and \ref{EA}, defined as the most general semantics of a feature which must apply in order for a particular \textit{norm} to be challenged. 
        
        Given that the RAIN Graph have \emph{coverage} in a particular socio-technical domain, the set of \textit{features} the Graph relates to in this domain can be seen as a guide to which parts of the context that are relevant. This helps to reduce the otherwise nebulous concept of a context into a more narrow form. With regards to the RAIN Graph, the context is whether the \textit{features} are present in the socio-technical sphere of the application or not. 
        
        The socio-technical use case of a project and who the stakeholders are are necessarily intertwined. Similar to how merging additional policy into a single RAIN Graph, representing the relationships between use cases, Stakeholders and Features will also naturally overlap and converge creating a reusable structure helping with knowledge elicitation and transfer.
        
        Features described readily lend themselves to ontology representation and dynamic questionnaires and dialogue approaches can be used to extract the details of an application without placing unduly high demands of expertise on the people characterizing the system. 
        
        \emph{Ex. The features Remote Processing, Personal Data, Anthropomorphic Human Interaction, Language Dependence, Vulnerable End Users, and Hazardous Robotics (stove) are present in the example home automation system. End Users, Developers, Procurers and Auditors are relevant stakeholders. These identified in the context layer of the pipeline.}
    
        \subsubsection{Assessment layer}
    
        Given that a certain set of \textit{features} and \textit{stakeholders} have been asserted by the context layer, some of the contexts of G will be active, and their statements will apply.
        The purpose of the assessment layer is to see if the rules (7) with regards to these contextualized \textit{norms} have been violated or not. Each of these assessment statements are connected to an appropriate type of test (e.g quiz, monitoring, supplied evidence).
        
        In this manner, no assessment is required in the cases where the context does not apply thereby preventing a bloat of irrelevant assessment questions. Every assessment test that do apply can be constructed towards a particular feature and stakeholder rather than towards the high-level goals or attempted generalisations. Because each negative assessment result violate a particular norm, and if this norm \textit{counts as} the high-level value, then the violations of the assessment rules will also be violations of the high-level norms they connect to.  
    
        \emph{Ex: RAIN assessment find that Remote Processing is used without a use-case reason (it is used to collect marketing data). Security measures surround handling of the stove and support exists for multiple languages. Anthropomorphic language is a \emph{Transparency} concern especially due to the Vulnerable End User feature.}.
    
        \subsubsection{Results layer and projections}

        When assessments have been performed, the result can be evaluated in several ways. A straightforward way is to enumerate the Values in set $V$ and determine what level of violation and thus maturity score which applies to each Value. This would be a RAIN Graph-specific result. Another straightforward way is to look to the context of a particular policy and similarly enumerate its particular HLVs together with the aggregated maturity level.
        A less straightforward but highly effective way is to provide a set of statements on the contents of $G$, where each statement maps to a particular requirement of an external assessment \textit{covered} by $G$.
        For instance if a RAIN Graph \textit{covers} GTAI, a set of statements on the graph can map the results to each of the assessment questions in the guidelines. This way a particular high-level policy can be assessed in a context-aware manner even if the policy itself is not constructed for the RAIN framework.
        
        Given the structure embedded in the RAIN graph, results can also be aggregated on particular stakeholders or socio-technical features, giving a valuable and detailed description on how ethical compliance is distributed over the socio-technical landscape of the Application.    
        
        \textit{Ex. While the system get high maturity levels on national safety standards, the aggregated GTAI scores are strongly violated due to the \emph{Remote Processing}, especially with regards to Privacy. The local Procurers internal guidelines are also found violated and the system is rejected. The developer of the system could adapt for on-site processing of recorded data to gain a higher \emph{Privacy} maturity level. Such adaption is a concrete technical and business problem, not an abstract ethical concern. After a switch to local processing, a less anthropomorphic language-use might further raise maturity level. Here the combined interests of all actors contribute towards an application with features in line with applied policies, driving towards more ethically full-featured applications promoting sustainable and responsible development. }
        
\section{Discussion and Future Work} 

        In this paper, we presented the RAIN framework; a structured methodology for translating high-level policy to concrete normative requirements and features. Using the \textit{count-as} operator, we can formally represent the socio-technical contexts where a policy is relevant for an application. Formal representation of value-context relations allows us to trace requirements and features to the values they represent in both a \textit{verifiable} and \textit{transparent} way. The framework allows a structured discussion and communication about AI systems in a low-overhead manner; enabling effective policy making, compliance checking and ethics in a concrete way. RAIN considers all levels of the emerging ecosystem of stakeholders: developers, procurers, users, regulators, and policymakers. 

        The RAIN Graph shifts the guidelines and assessment criteria from abstract values to contextualised features and requirements. In contrast to high-level AI Ethics, software development is already apt at working with such feature requirements. Expert knowledge embedded in the graph decreases the overhead of local practical-philosophy and policy expertise and complicated organisational containment-strategies, thus increasing both the availability and impact of any policy. 
        
        As more AI products are marketed, complex software containing multiple AI modules developed by multiple developers procured by yet other public or commercial organisations will become more common. Applying a RAIN Graph based assessment from the module level up, and from the top-organisational level down facilitates full-chain modular policy compliance checking and charting of responsibility. Cross-application of local organisational policies and national and international guidelines allows procurers to set their own terms and requirements on their suppliers, enabling each layer of the chain to take responsibility \cite{rubenstein2021acquiring,falco2021governing}. This structured approach applied on a top-policymaking level enables top-down discussions focused on socio-technical hot-spots rather than nebulous and hard-to-define AI.

    Continuing on alleviating the overhead on the AI ecosystem, our future work includes adding a functional model of AI systems to the graph representation which would extend the scope from high-level principles to a more direct multi-level treatment of explainability, contestability, and trust. Finally, our future work also includes field testing of tools and methodologies building on the presented framework.
    
\begin{acknowledgments}
  This work was supported by the Wallenberg AI, Autonomous Systems and Software Program (WASP) funded by the Knut and Alice Wallenberg Foundation. Brännström, Theodorou and Dignum thank the Knut and Alice Wallenberg Foundation for grant RAIN (2020:2012) that supported their efforts.  
\end{acknowledgments}

\bibliography{rain}

\appendix

\end{document}